\documentclass[12pt]{article}
\usepackage[latin1]{inputenc}
\usepackage{amssymb}
\usepackage{graphicx}
\usepackage{color}
\newcommand{\dt}{\partial_t}
\newcommand{\dx}{\partial_x}
\newcommand{\dy}{\partial_y}
\newcommand{\dz}{\partial_z}

\newcommand{\eps}{\varepsilon}
\newcommand{\dsp}{\displaystyle}
\newcommand{\R}{{\mathbb R}}

\newcommand{\sech}{\mbox{\textnormal{sech}}}

\newcommand{\cG}{{\mathcal G}}
\newcommand{\cQ}{{\mathcal Q}}

\newtheorem{rema}{Remark}[section]{\bf}{\rm}
\begin{document}
\title{Derivation of asymptotic two-dimensional time-dependent equations for ocean wave propagation}
\author{David Lannes$^{(1)}$ and Philippe Bonneton$^{(2)}$  \\ 
\small (1) Université Bordeaux 1; IMB and CNRS UMR 5251, Talence, F-33405 France\\ 
\small (2) Université Bordeaux 1; CNRS; UMR 5805-EPOC, Talence, F-33405 France \\ 
\small (e-mail: David.Lannes@math.u-bordeaux1.fr; p.bonneton@epoc.u-bordeaux1.fr )}
\date{\ }
\maketitle
\begin{abstract}
A general method for the derivation of asymptotic nonlinear shallow water and deep water models is presented. 
Starting from a general dimensionless version of the water-wave equations, we reduce the problem to a system of two
equations on the surface elevation and the velocity potential at the
free surface. These equations involve a Dirichlet-Neumann operator and we 
show that all the asymptotic models can be recovered by a simple asymptotic expansion of this operator, in function of the shallowness parameter (shallow water limit) or the steepness parameter (deep water limit). Based on this method, a  new two-dimensional fully dispersive model for small wave steepness is also derived, which extends to uneven bottom the approach developed by Matsuno \cite{matsuno3} and Choi \cite{choi}. This model is still valid in shallow water but with less precision than what can be achieved with  Green-Naghdi model, when fully nonlinear waves are considered. The combination, or the coupling, of the new fully dispersive equations with the fully nonlinear shallow water Green-Naghdi equations represents a relevant model for describing ocean wave propagation from deep to shallow waters.
\end{abstract}
\newpage
%
%
\section{Introduction}
The propagation of surface waves through an incompressible homogenous inviscid fluid is described by the 3D Euler equations combined with nonlinear boundary conditions at the free-surface and at the bottom. This problem is extremely difficult to solve, in particular because the moving surface boundary is part of the solution. The complexity of this problem led physicists, oceanographers and mathematicians to derive simpler sets of equations in some specific physical regimes. Equations thus obtained may be divided into two groups, namely shallow-water models and deep-water models. One of the goals of this
paper is to clarify the range of validity of these models and to show
why at least two different asymptotic models are necessary for a correct description of ocean waves.

In shallow water conditions, the classical approach is based on a perturbation method with respect to a small parameter $\mu=(h_0/L)^2$ ($h_0$ the characteristic water depth, $L$ the characteristic horizontal scale), in order to reduce the three-dimensional equation system to a two-dimensional one. This method, initially introduced by Boussinesq \cite{bou1872}, allows to derive  several  shallow-water equations, which are named "Boussinesq-type equations". A large class of such equations can be expressed in the following 2D nondimensional form:
\begin{equation}\label{eqBouType}
	\left\lbrace
	\begin{array}{l}
	\dt\zeta+\nabla\cdot(hV)=O(\mu^2) \label{eqbou}\\
	\dt V +\eps (V\cdot\nabla)V+\nabla\zeta=\mu {\cal D}+O(\mu^2)
	\end{array}\right.
\end{equation}
where $\epsilon=a/h_0$ ($a$ the order of free surface amplitude) is the nonlinearity parameter and the dimensionless flow variables are: $h$ the water depth, $\zeta$  the surface elevation and $V$ the depth-averaged velocity. $\cal D$ characterizes non-hydrostatic and dispersive effects and is a function of  wave variables and their derivatives. Higher-order Boussinesq equations can be derived (e.g. Madsen et Schäffer \cite{mad1999}), but in this paper we restrict our analysis to $O(\mu^2)$, which is a good approximation for most of nearshore wave applications
(e.g. Madsen et al. \cite{mad1997a}, \cite{mad1997b}, Cienfuegos et al. \cite{cie2007}).

 In its classical form, Boussinesq wave theory is a one-dimensional approach based on the assumptions of weak dispersion, weak non-linearity and balance between dispersion and non-linearity: $\epsilon=O(\mu)\ll 1$. The KdV \cite{kdv} and BBM \cite{bbm} equations can also be derived in the case of unidirectional waves; this approach is relevant to study fundamental wave dynamics problems, such as 1D solitary wave propagation on flat bottom (e.g. Boussinesq \cite{bou1872} and Benjamin \cite{benjamin}). However, the classical Boussinesq assumptions may severely restrict applicability to real word wave propagation problems. Applications to coastal zone, have motivated theoretical developments (see reviews by Dingemans \cite{din1997}, Madsen et Schäffer \cite{mad1999}, Kirby \cite{kir2003} and Barthélemy \cite{bar2004}) for extending the range of applicability of Boussinesq-type equations in term of
varying bottom, dispersive  and  nonlinearity effects, which play an important role in the nearshore wave dynamics.   

The 2D Boussinesq equations ($\epsilon=O(\mu)\ll 1$) for   non flat bottom were first derived by Peregrine \cite{per1967}:
\begin{equation}\label{eqPeregrine}
{\cal D}=\frac{h}{2}\nabla(\nabla\cdot(h\dt V))-\frac{h^2}{6}\nabla^2\dt V \ \label{eqpere}.
\end{equation}
For many coastal applications the weak dispersion of these equations is a critical limitation. Witting \cite{wit1984} proposed a method based on Padé expansion of the exact linear phase velocity to improve Boussinesq-type equations. From this method, several equations (order $O(\mu^2)$) with improved dispersion characteristic have been derived (e.g. Madsen et al. \cite{mad1991}, Nwogu \cite{nwo1993}, Schäffer and Madsen \cite{sch1995}, Bona et al. \cite{BCS,BCL}).

In 1953, a breakthrough treating nonlinearity was made by Serre (see Barthélemy \cite{bar2004} for a review). He derived 1D fully nonlinear ($\epsilon=O(1)$) weakly dispersive equations for horizontal bottom:
\begin{equation}\label{eqSerre1}
{\cal D}=\frac{1}{3h}\dx (h^3(V_{xt}+VV_{xx}-(V_{x})^2) ) .
\end{equation}
The same system was obtained later by Su and Gardner \cite{sug1969}. Seabra-Santos et al. \cite{sea1987} have provided an extension of this model to non flat bottom:
\begin{equation}\label{eqSerre2}
{\cal D}=-\frac{1}{h}\left( \dx (h^2(\frac{1}{3}P+\frac{1}{2}Q))+b_xh(\frac{1}{2}P+Q)  \right) ,
\end{equation}
with $P=-h(V_{xt}+VV_{xx}-(V_{x})^2)$ and $Q=b_x(V_{t}+VV_{x}+b_{xx}V^2)$. Dingemans \cite{din1997} expressed mistrust for the validity of these equations for uneven topographies, considering that the derivation required the assumption of vertical uniformity of the horizontal velocity. Cienfuegos et al. \cite{cie2006} have shown that Serre equations can be formally established without assumption on the velocity profile.

Finally, Green and Naghdi \cite{gre1976} derived 2D fully nonlinear  weakly dispersive equations for uneven bottom which represents a two-dimensional extension of Serre equations. Except for being formulated in terms of the velocity vector at an arbitrary $z$ level, the equations of Wei et al. \cite{wei1995} are basically equivalent to the 2D Serre or Green-Naghdi equations; this is also
the case of the equations derived in \cite{MS} through Hamilton's principle.

The range of validity of all the models introduced 
above may vary as far as the nonlinearity parameter $\eps$
is concerned, but they all require that the shallowness parameter $\mu$
is small. In deep water ($\mu \not\ll 1$) the Boussinesq wave theory fails
but it is yet possible to derive asymptotic expansions from the water-wave
equations under the condition that the \emph{steepness} $\eps\sqrt{\mu}=a/\lambda$ is small. Such an approach was first made in 1D and flat
bottoms by Matsuno \cite{matsuno1} and extended to uneven bottoms \cite{matsuno2}, 2D weakly transverse
waves \cite{matsuno3} and higher order expansions \cite{choi}. 
Since one always has $\eps\lesssim 1$, the
small steepness assumption $\eps\sqrt{\mu}\ll 1$ is also satisfied
in the shallow water regime $\mu\ll 1$ discussed above and this is the
reason why it is often claimed that the models derived
in \cite{matsuno1,matsuno2,matsuno3,choi} are valid in the whole
range $\mu\in(0,\infty)$. However, as we show here, their precision
is then far below the one of the Green-Naghdi equations.

In this paper we propose a systematic derivation of all the models evoked
in this introduction (and of some new ones); to this end, we use the global
method introduced in the recent mathematical work \cite{alvarezlannes}. 
Starting from a general nondimensionalized version of the water-wave equations (which takes into accounts the different nondimensionalizations used
in deep and shallow water), following \cite{Zakharov,CSS}, we reduce the problem to a system of two
equations on the surface elevation and the velocity potential at the
free surface. These equations involve a Dirichlet-Neumann operator and we 
show that
all the asymptotic models can be recovered by a simple asymptotic expansion
of this operator. In the deep water setting ($\mu\not\ll 1$) this expansion
is made with respect to $\eps$ (which is small, as a consequence of 
the small steepness assumption $\eps\sqrt{\mu}\ll 1$ and the deep water assumption $\mu\not\ll 1$); in the shallow-water
setting, we rather choose to expand the Dirichlet-Neumann  operator in
terms of $\mu$ since $\eps$ is not always small (e.g., $\eps\sim 1$ for Green-Naghdi).

As said above, all the asymptotic models are obtained in a systematic way
with the present approach. In addition, we derive a new fully dispersive model for 3D waves over uneven bottom.
Our method brings clarifications about the validity domain 
of the main asymptotic wave models. In particular,
\begin{itemize}
	\item We can easily show that the shallow
	water limit of the Dirichlet-Neumann expansion
	corresponding to the deep-water models has a very poor 
	precision in the fully nonlinear case $\eps\sim 1$;
	this proves that the deep-water models such as those 
	of \cite{matsuno1,matsuno2,matsuno3,choi} are not in general
	suitable for nonlinear shallow-water waves;
	\item We do not make any assumption on the velocity profile 
	or on any related quantity.
	The only assumptions we make concern the value 
	of the parameters $\eps$ and 
	$\mu$ (and a third parameter $\beta$ linked to 
	the amplitude of the
	bottom variations for uneven bottoms); the properties of the 
	velocity profile (and in particular its vertical behavior) 
	are then rigorously established; we
thus believe that this article should clarify the discussions concerning
the assumptions made on the velocity profile, and in particular the controversy
raised in \cite{din1997} about the Serre equations.
\end{itemize}

%
%
\section{General nondimensionalized water-wave equations}
\label{appnd}
Parameterizing the free surface by $z=\zeta(t,X)$ (with $X=(x,y)\in\R^2$) 
and the bottom by
$z=-h_0+b(X)$ (with $h_0>0$ constant), one can use the incompressibility and 
irrotationality conditions to
write the water-wave equations under Bernouilli's formulation, in terms
of a  potential velocity $\phi$ (i.e., the velocity field is given by 
${\mathbf v}=\nabla_{X,z}\phi$):
\begin{equation}
	\label{eqbern}
	\left\lbrace
	\begin{array}{lcl}
	\displaystyle \partial_{x }^2\phi 
	+\partial_{y }^2\phi 
	+\partial_{z }^2\phi =0,
	& & -h_0+b\leq z \leq \zeta ,\\
	\displaystyle \partial_n\phi=0, & &z =-h_0+b ,\\
	\displaystyle \partial_{t }\zeta +\nabla\zeta\cdot\nabla\phi
	=\partial_{z }\phi,
	& &  z = \zeta ,\\
	\displaystyle \partial_{t }\phi 
	+\frac{1}{2}\big(\vert\nabla\phi\vert^2+
	(\partial_{z }\phi )^2\big)+g\zeta =0,
	& & z =\zeta,
	\end{array}
	\right.
\end{equation}
where $g$ is the gravitational acceleration, $\nabla=(\dx,\dy)^T$ and $\partial_n\phi$ is the outward normal
derivative at the boundary of the fluid domain.

In deriving approximate equations by asymptotic methods it is necessary to introduce dimensionless quantities based on characteristic scales for the wave motion. Four main length scales are involved in this problem: $h_0$ the characteristic water depth, $L$ the characteristic horizontal scale, $a$ the order of free surface amplitude and $B$ the order of bottom topography variation. Three independant dimensionless parameters can be formed from these four scales. We choose:
\begin{equation}
\frac{a}{h_0}=\epsilon, \  \  \ \frac{h_0^2}{L^2}=\mu, \ \ \  \frac{B}{h_0}=\beta,
\end{equation}
where $\epsilon$ is often called the nonlinearity parameter, while $\mu$ is the shallowness parameter. 

Commonly two distinct nondimensionalizations are used in oceanography (e.g. Dingemans (1997)) depending on the value of $\mu$: shallow water scaling ($\mu \ll 1$) and Stokes wave scaling for intermediate to deep water ($\mu \not\ll 1$). In this paper, we present a general nondimensionalization which applies to any wave regime.
The order of magnitude of wave motion variables are given by the linear wave theory. In particular, $(gh_0\nu)^{1/2}$ and $\frac{aL}{h_0}(\frac{gh_0}{\nu})^{1/2}$ are the characteristic scales of respectively the wave celerity and the potential velocity, with $\nu=\tanh(\mu^{1/2})/\mu^{1/2}$.

Let us normalize all variables according to the scales anticipated on physical grounds:
$$
	\begin{array}{llll}
	x=L x',&
	y=L y',&
	z=h_0\nu z',&
	t=\frac{\lambda}{\sqrt{gh_0\nu}}t',\\
	\zeta=a\zeta',&
	\Phi=\frac{a}{h_0}L\sqrt{\frac{gh_0}{\nu}}\Phi',&
	b=Bb'.
	\end{array}
$$
From this general nondimensionalization we can recover the classical scalings for shallow and deep water: $\nu \sim 1$ when $\mu \ll 1$  and $\nu \sim \mu^{-1/2}$ when $\mu \gg 1$. 

The equations of motion (\ref{eqbern}) then become (after dropping the
primes for the sake of clarity):
\begin{equation}
	\label{app1}
	\left\lbrace
	\begin{array}{l}
	\displaystyle \nu^2\mu\partial_{x }^2\Phi 
	+\nu^2\mu\partial_{y }^2\Phi 
	+\partial_{z }^2\Phi =0,
	\qquad \frac{1}{\nu}(-1+\beta b )\leq z \leq \frac{\eps}{\nu}\zeta ,\vspace{1mm}\\
	\displaystyle -\nu^2\mu\nabla(\frac{\beta}{\nu}b )\cdot\nabla\Phi +\partial_{z }\Phi =0,
        \qquad z =\frac{1}{\nu}(-1+\beta b ),\vspace{1mm}\\
	\displaystyle \partial_{t }\zeta -\frac{1}{\mu\nu^2}\big(-\nu^2\mu\nabla(\frac{\eps}{\nu}\zeta )\cdot\nabla\Phi
	+\partial_{z }\Phi \big)
	=0,
	\qquad  z = \frac{\eps}{\nu}\zeta ,\vspace{1mm}\\
	\displaystyle \partial_{t}\Phi+\frac{1}{2}\big(\frac{\eps}{\nu}\vert\nabla\Phi\vert^2
        +\frac{\eps}{\mu\nu^3}(\partial_{z }\Phi )^2\big)+\zeta =0,
	\qquad z =\frac{\eps}{\nu}\zeta .
	\end{array}
	\right.
\end{equation}

In order to reduce this set of equations into a system of two evolution
equations, we introduce the trace of the velocity potential at the free
surface, namely $\psi=\Phi_{\vert_{z=\eps/\nu\zeta}}$ and the Dirichlet-Neumann operator 
$\cG^\nu_{\mu}[\frac{\eps}{\nu}\zeta]\cdot$ as
$$
  \cG^\nu_{\mu}[\frac{\eps}{\nu}\zeta]\psi
	=-\nu^2\mu\nabla(\frac{\eps}{\nu}\zeta )\cdot\nabla\Phi_{\vert_{z=\frac{\eps}{\nu}\zeta}}
	+\partial_{z }\Phi_{\vert_{z=\frac{\eps}{\nu}\zeta}},
$$
with $\Phi$ solving the boundary value problem
$$
	\left\lbrace
	\begin{array}{l}
	\dsp \nu^2\mu\partial_{x }^2\Phi 
	+\nu^2\mu\partial_{y }^2\Phi 
	+\partial_{z }^2\Phi =0,
	\qquad \frac{1}{\nu}(-1+\beta b )\leq z \leq \frac{\eps}{\nu}\zeta,
	\vspace{1mm}\\
	\dsp \Phi_{\vert_{z=\frac{\eps}{\nu}\zeta}}=\psi,\qquad
	\partial_n\Phi_{\vert_{z=\frac{1}{\nu}(-1+\beta b)}}=0,	
	\end{array}\right.
$$
(one can check that $\cG^\nu_{\mu}[\frac{\eps}{\nu}\zeta]\psi=\sqrt{1+\vert\nabla(\frac{\eps}{\nu}\zeta)\vert^2}\partial_n\Phi_{\vert_{z=\frac{\eps}{\nu}\zeta}}$, where $\partial_n\Phi$ stands for the upwards nondimensionalized normal
derivative at the surface). As remarked in
\cite{Zakharov,CSS}, the equations (\ref{app1}) are
equivalent to a set of two equations on the free surface parameterization
$\zeta$ and the trace of the velocity potential at the surface $\psi=\Phi_{\vert_{z=\eps/\nu\zeta}}$ involving the Dirichlet-Neumann operator. Namely,
\begin{equation}
	\label{app2}
	\left\lbrace
	\begin{array}{l}
	\dsp \dt \zeta-\frac{1}{\mu\nu^2}\cG^\nu_{\mu}[\frac{\eps}{\nu}\zeta]\psi=0,\\	
	\dsp \dt\psi+\zeta+\frac{\eps}{2\nu}\vert\nabla\psi\vert^2-
        \frac{\eps\mu}{\nu^3}
        \frac{(\frac{1}{\mu}\cG^\nu_{\mu}[\frac{\eps}{\nu}\zeta]\psi+\nu\nabla(\eps\zeta)\cdot\nabla\psi)^2}
        {2(1+\eps^2\mu\vert\nabla\zeta\vert^2)}=0.
	\end{array}\right.
\end{equation}
In order to derive the system (\ref{nondimww}), let $\cG_{\mu}[\eps\zeta,\beta b]\cdot$ be
the Dirichlet-Neumann operator $\cG_{\mu}^\nu[\frac{\eps}{\nu}\zeta]\cdot$ corresponding to the case $\nu=1$. One will easily check that
$$
	\forall\nu>0,\qquad
	\cG_{\mu}[\eps\zeta,\beta b]
	=\frac{1}{\nu}\cG_{\mu}^\nu[\frac{\eps}{\nu}\zeta],
$$
so that plugging this relation into (\ref{app2}) yields
\begin{equation}\label{nondimww}
	\left\lbrace
	\begin{array}{l}
	\dsp \dt \zeta-\frac{1}{\mu\nu}\cG_{\mu}[\eps\zeta,\beta b]\psi=0,\\
	\dsp \dt \psi+\zeta+\frac{\eps}{2\nu}\vert\nabla\psi\vert^2-\frac{\eps\mu}{\nu}
        \frac{(\frac{1}{\mu}\cG_{\mu}[\eps\zeta,\beta b]\psi+\nabla(\eps\zeta)\cdot\nabla\psi)^2}
        {2(1+\eps^2\mu\vert\nabla\zeta\vert^2)}=0.
	\end{array}\right.
\end{equation}
The system (\ref{nondimww}) recasts the water-wave equations in terms
of the free surface elevation $\zeta$ and the velocity potential $\psi$
at the surface; this is the formulation which will serve as a basis for all
the computations in this article.

%
%
\section{Shallow water models}

This section is devoted to the study of shallow water waves: $\mu\ll 1$.
The most general situation (when no assumption is made on $\eps$ and $\beta$)
is addressed
in \S \ref{sectSerre}, where the
the Green-Naghdi
equations (also called Serre \cite{ser1953} or \emph{fully} nonlinear Boussinesq equations
 \cite{wei1995}) are derived. The key point is an asymptotical
expansion of the Dirichlet-Neumann operator $\cG_\mu[\eps\zeta]$ 
with respect to $\mu$.\\
We then briefly show how to recover simpler models under additional
assumptions on $\eps$; the \emph{moderately} nonlinear case $\eps=O(\sqrt{\mu})$
is addressed in \S \ref{sectmild} and the \emph{weakly} nonlinear case 
$\eps=O(\mu)$ in \S \ref{sectweak}.
Some new models are then derived in the next sections.

\subsection{The fully nonlinear case:  $\epsilon\sim1$, $\beta\sim1$ and $\mu\ll1$ }\label{sectSerre}
From the assumption $\mu\ll 1$, one gets $\nu\sim1$ and we take $\nu=1$
in (\ref{nondimww})
to simplify (this corresponds to the usual 
shallow-water nondimensionalization). 

\subsubsection{Asymptotic expansion of the Dirichlet-Neumann operator}

Since $\mu\ll1$, we look for an asymptotic expansion of $\Phi$ under the form
\begin{equation}\label{n0}
	\Phi_{app}=\sum_{j=0}^N\mu^j\Phi_j.
\end{equation}
Plugging this expression into the nondimensionalized water-wave equations one
can cancel the residual up to the order $O(\mu^{N+1})$ provided that
\begin{equation}\label{n1}
	\forall j=0,\dots,N,\qquad
	\dz^2\Phi_{j}=-\dx^2\Phi_{j-1}-\dy^2\Phi_{j-1}
\end{equation}
(with the convention that $\Phi_{-1}=0$), together with the boundary
conditions
\begin{equation}\label{n2}
	\forall j=0,\dots,N,\qquad
	\left\lbrace
	\begin{array}{l}
	\Phi_{j}\,_{\vert_{z=\eps\zeta}}=\delta_{0,j}\psi,\\	
	\dz \Phi_j=\beta\nabla b\cdot \nabla\Phi_{j-1}\,_{\vert_{z=-1+\beta b}}
	\end{array}\right.
\end{equation}
(where $\delta_{0,j}=1$ if $j=0$ and $0$ otherwise).\\
Solving the ODE (\ref{n1}) with (\ref{n2}) is completely straightforward, 
and this procedure can be implemented on any symbolic computation 
software to compute the $\Phi_j$ at any order. For our purposes here, 
we must take $N=1$ and thus need to compute $\Phi_0$ and $\Phi_1$; one finds
\begin{eqnarray}
	\label{n4} \Phi_0&=&\psi,\\
	\label{n5} \Phi_1&=&(z-\eps\zeta)(-\frac{1}{2}(z+\eps\zeta)-1+\beta b)\Delta \psi+\beta(z-\eps\zeta)\nabla b\cdot\nabla\psi.
\end{eqnarray}
\begin{rema}
	This shows that at order $O(\mu^2)$, and without assumption on $\eps$,
	the potential depends quadratically on $z$, as is well-known.
	When $\mu$ is not small, the expansion (\ref{n0}) is not convergent
	and this is the reason why we use another technique to expand
	the Dirichlet-Neumann operator in \S \ref{sectmatsuno}.
\end{rema}

An approximation of order $O(\mu^{N+1})$ of $\cG_{\mu}[\eps\zeta,\beta b]\psi$
is then given by
\begin{equation}\label{n3}
	\cG_{\mu}[\eps\zeta,\beta b]\psi=-\mu\nabla(\eps\zeta )\cdot\nabla\Phi_{app}\,_{\vert_{z=\eps\zeta}}
	+\partial_{z }\Phi_{app}\,_{\vert_{z=\eps\zeta}}+O(\mu^{N+2});
\end{equation}
for $N=1$, we thus obtain from (\ref{n0}), (\ref{n4}) and (\ref{n5}),
\begin{equation}\label{expDN}
	\cG_{\mu}[\eps\zeta,\beta b]\psi=-\mu \nabla\cdot (h\nabla\psi)+
	\mu^2\nabla\cdot{\mathcal T}[h,\beta b]\cdot\nabla\psi+O(\mu^3)
\end{equation}
where $h=1+\eps\zeta-\beta b$ and where the linear operator 
${\mathcal T}[h,\beta b]$ is defined as 
\begin{equation}\label{eqT0}
	{\mathcal T}[h,\beta b]\cdot W=
	{\mathcal T}_{1}^* h	{\mathcal T}_{1}\cdot W+
	{\mathcal T}_{2}^* h	{\mathcal T}_{2}\cdot W.
\end{equation}
with ${\mathcal T}_j^*$ ($j=1,2$) denoting the adjoint of the operators
${\mathcal T}_j$ given by
\begin{eqnarray}
	\label{eqT1}
	{\mathcal T}_{1}\cdot W:=\frac{h}{\sqrt{3}}\nabla\cdot W-\frac{\sqrt{3}}{2}\beta\nabla b\cdot W,
	\quad \mbox{ and }\quad
	{\mathcal T}_{2}\cdot W:=\frac{1}{2}\beta\nabla b\cdot W. 
\end{eqnarray}

\subsubsection{Derivation of the Green-Naghdi equations}

We use here the asymptotic expansion (\ref{expDN}) to derive the
Green-Naghdi equations (also called  Serre equations) from the full water waves
equations (\ref{nondimww}).

One can equivalently write (\ref{expDN}) under the form
\begin{equation}\label{n6}
	\cG_{\mu}[\eps\zeta,\beta b]\psi=
	-\mu \nabla\cdot(h {\bf v})+O(\mu^3),
\end{equation}
where ${\bf v}=\nabla\psi-\mu\frac{1}{h}{\mathcal T}[h,\beta b] \nabla\psi$.
\begin{rema}
	According to formulae (\ref{n4})-(\ref{n5}), the horizontal component
	of the velocity in the fluid domain is given by 
	$V(z)=\nabla\Phi_0+\nabla\Phi_1(z)+O(\mu^2)$, and one can check
	that 
	${\bf v}=\frac{1}{h}\int_{-1+\beta b}^{\eps \zeta}V(z)dz+O(\mu^2)$,
	so that, in accordance with the classical derivations of the
	Green-Naghdi equations, \emph{${\bf v}$ is the vertically averaged
	horizontal component of the velocity} (up to $O(\mu^2)$ terms).
\end{rema}

Recalling that $\nu=1$ here,  it follows from (\ref{n6}) that the first
equation of (\ref{nondimww}) can be written
$\dt \zeta+\nabla\cdot(h{\bf v})=O(\mu^2)$.\\
From the definition of ${\bf v}$ given in (\ref{n6}), one gets easily
$$
	\nabla\psi={\bf v}+\mu\frac{1}{h}{\mathcal T}[h,\beta b]{\bf v}+O(\mu^2);
$$
taking the gradient of the second equation of (\ref{nondimww}), replacing
$\nabla\psi$ in the resulting equation by the above formula, 
using (\ref{n6}), and neglecting the $O(\mu^2)$ quantities gives  
the Green-Naghdi
equations (recall that $h=1+\eps\zeta-\beta b$):
\begin{equation}\label{eqGN}
	\left\lbrace
	\begin{array}{l}
	\displaystyle \dt\zeta+\nabla\cdot(h{\bf v})=0,\vspace{1mm}\\

	\displaystyle(1+\frac{\mu}{h}{\mathcal T}[h,\beta b])\dt {\bf v}+\nabla\zeta
	+\eps ({\bf v}\cdot\nabla){\bf v}\\
	\displaystyle\indent+\mu\eps
	\Big\lbrace-\frac{1}{3h}\nabla\big[h^3\big(({\bf v}\cdot \nabla) (\nabla\cdot {\bf v})-(\nabla\cdot {\bf v})^2\big)\big]+\cQ[h,\beta b]({\bf v})\Big\rbrace=0,
	\end{array}\right.
\end{equation}
where 
\begin{eqnarray}
{\mathcal T}[h,\beta b]\cdot W&=&{\mathcal T}_{1}^* h	{\mathcal T}_{1}\cdot W+{\mathcal T}_{2}^* h	{\mathcal T}_{2}\cdot W \\ &=&-\frac{1}{3}\nabla(h^3\nabla\cdot W)+\beta \frac{1}{2}\big[\nabla(h^2\nabla b \cdot W)-h^2\nabla b \nabla\cdot W\big]+\beta^2 h\nabla b\nabla b\cdot W  \nonumber
\end{eqnarray}
and the purely topographical term 
$\cQ[h,\beta b]({\bf v})$ (which is quadratic in ${\bf v}$) is
defined as
\begin{eqnarray}
	\nonumber
	\cQ[h,\beta b]({\bf v})
	&=&\frac{\beta}{2h}\big[\nabla\big(h^2({\bf v}\cdot\nabla)^2b\big)
	-h^2\big(({\bf v}\cdot \nabla) (\nabla\cdot {\bf v})
	-(\nabla\cdot {\bf v})^2\big)\nabla b\big]\\
	& &+\beta^2\big(({\bf v}\cdot\nabla)^2b\big)\nabla b.
\end{eqnarray}
\begin{rema}\label{remreg}
	The formulation (\ref{eqGN}) of the Green-Naghdi equation
	is not at first sight the same as usual
	(see \cite{MS} and \cite{wei1995}). 
	A closer look shows however that
	they are exactly the same, as expected. The interest of the
	present form is that the second equation gives a straightforward
	control of the quantities $(h{\bf v},{\bf v})$, $(h{\mathcal T}_1{\bf v},{\mathcal T}_1{\bf v})$ and $(h{\mathcal T}_2{\bf v},{\mathcal T}_2{\bf v})$ (with ${\mathcal T}_j$ given by (\ref{eqT1})). This control yields
regularizing effects of the same kind as those of the BBM equation compared
to KdV, and is thus expected to ease numerical computations (work in progress).
\end{rema}

In \cite{wei1995}, Wei et al. derived some Green-Naghdi
equations (fully nonlinear Boussinesq models in that reference) with
improved frequency dispersion by replacing the vertically averaged horizontal
velocity ${\bf v}$ by the velocity ${\bf v}_\alpha$ taken at some intermediate
depth $z_\alpha(x,y)$ (thus following the approach of Nwogu \cite{nwo1993} for
weakly nonlinear Boussinesq systems). Such systems could of course be
derived similarly from (\ref{eqGN}).

\subsection{The moderately nonlinear case: 
$\mu\ll1$ and $\eps=O(\sqrt{\mu})$}\label{sectmild}

The simplifications that can be made on the Green-Naghdi equations
(\ref{eqGN}) under the assumption $\eps=O(\sqrt{\mu})$ 
depend on the topography. As for the surface
variations, we distinguish three different regimes:\\
{\it (a) Fully nonlinear topography: $\beta=O(1)$.} In this
case, no significative simplification can be made, and the
full equations must be kept.\\
{\it (b) Moderately nonlinear topography: $\beta=O(\sqrt{\mu})$.} In this
regime, the last term of the second equation of (\ref{eqGN})
can be written
$$
	-\frac{\mu\eps}{3}\nabla\big[\big(({\bf v}\cdot \nabla) 
	(\nabla\cdot {\bf v})-(\nabla\cdot {\bf v})^2\big)\big]
	+O(\mu^2),
$$
so that one can replace (\ref{eqGN}) by
$$
	\left\lbrace
	\begin{array}{l}
	\displaystyle \dt\zeta+\nabla\cdot(h{\bf v})=0,\vspace{1mm}\\
	\displaystyle(1+\frac{\mu}{h}{\mathcal T}[h,\beta b])\dt {\bf v}+\nabla\zeta
	+\eps ({\bf v}\cdot\nabla){\bf v}
	-\frac{\mu\eps}{3}\nabla\big[\big(({\bf v}\cdot \nabla) (\nabla\cdot {\bf v})-(\nabla\cdot {\bf v})^2\big)\big]=0,
	\end{array}\right.
$$
(note that some simplifications could also be made in the term $\frac{\mu}{h}{\mathcal T}[h,\beta b]\dt {\bf v}$ but they are not interesting because
they would partially destroy the regularizing effects evoked in Remark
\ref{remreg}).

{\it (c) Weakly nonlinear topography: $\beta=O(\eps)$.} This stronger assumption
allows a simplification of the term  $\frac{\mu}{h}{\mathcal T}[h,\beta b]\dt {\bf v}$, and one gets
$$
	\left\lbrace
	\begin{array}{l}
	\displaystyle \dt\zeta+\nabla\cdot(h{\bf v})=0,\vspace{1mm}\\
	\displaystyle\big(1-\frac{\mu}{3h}\nabla (h^3 \nabla\cdot)\big)
	\dt {\bf v}+\nabla\zeta
	+\eps ({\bf v}\cdot\nabla){\bf v}
	-\frac{\mu\eps}{3}\nabla\big[\big(({\bf v}\cdot \nabla) (\nabla\cdot {\bf v})-(\nabla\cdot {\bf v})^2\big)\big]=0;
	\end{array}\right.
$$
the main interest of this model is that in the case of flat bottoms
($b=0$), its unidirectional limit gives the Cammassa-Holm equation
(\cite{CH,CHH,Johnson,CL}).

\subsection{The weakly nonlinear case: $\mu\ll 1$ and $\eps=O(\mu)$}
\label{sectweak}

The assumption $\eps=O(\mu)$ is the classical long waves assumption
which yields the usual Boussinesq models. Here again, we
briefly show how to recover these models in different topographic
regimes:\\
{\it (a) Fully nonlinear topography: $\beta=O(1)$.} Neglecting in (\ref{eqGN})
the terms which are of order $O(\mu^2)$ under the assumption $\eps=O(\mu)$
yields the equations
$$
	\left\lbrace
	\begin{array}{l}
	\displaystyle \dt\zeta+\nabla\cdot(h{\bf v})=0,\vspace{1mm}\\
	\displaystyle(1+\frac{\mu}{h}{\mathcal T}[h,\beta b])\dt {\bf v}+\nabla\zeta
	+\eps ({\bf v}\cdot\nabla){\bf v}=0,
	\end{array}\right.
$$
equivalent to equations (\ref{eqbou}) and (\ref{eqpere}) and which correspond to the Boussinesq model for nonflat bottoms derived by Peregrine
\cite{per1967}.

For many coastal applications the weak dispersion of these equations is a critical limitation. Several alternative equations, with the same approximations $\mu \ll 1$ and $\eps=O(\mu)$, have been developed to improve dispersion properties (e.g. Madsen et al. \cite{mad1991}, Nwogu \cite{nwo1993}, Schäffer and Madsen \cite{sch1995} or Bona et al. \cite{BCS,BCL}).

{\it (b) Weakly nonlinear topography: $\beta=O(\mu)$.} In this case, the
equations simplify further into
\begin{equation}\label{weakBouss}
	\left\lbrace
	\begin{array}{l}
	\displaystyle \dt\zeta+\nabla\cdot(h{\bf v})=0,\vspace{1mm}\\
	\displaystyle(1-\frac{\mu}{3}\Delta)\dt {\bf v}+\nabla\zeta
	+\eps ({\bf v}\cdot\nabla){\bf v}=0,
	\end{array}\right.
\end{equation}
for which topographic effects play a role only through the presence of 
$h=1+\eps\zeta-\beta b$ in the first equation. 
\begin{rema}\label{remrec}
	This latter Boussinesq system corresponds to the
	following approximation of the Dirichlet-Neumann operator:
	$$
		\cG_\mu[\eps\zeta,\beta b]\psi=
		-\mu \nabla\cdot (h\nabla \psi)
		-\frac{\mu^3}{3}\nabla\cdot \Delta\nabla\psi+O(\mu^3),
	$$
	which is deduced from  (\ref{n6}) by neglecting the terms
	which are $O(\mu^3)$ in the present scaling.
\end{rema}

\section{Fully dispersive models:  $\eps\sqrt{\mu}\ll 1$ and $\epsilon \ll 1$}
\label{sectmatsuno}

Small steepness asymptotics ($\eps\sqrt{\mu}\ll 1$) have been
introduced by Matsuno \cite{matsuno1} and generalized 
in \cite{matsuno2,matsuno3,choi}. It is often claimed 
that the derived equations are valid under the condition 
$\eps\sqrt{\mu}\ll1$ only,
and that the other asymptotic models (which satsify this condition)
can be deduced from them. We show here that this is not always the case.
The unified framework used in this article allows us to check
 for
instance that the Matsuno equations do not degenerate into 
the Green-Naghdi equations in shallow-water and that their
precision is much smaller.\\
After making a new asymptotic expansion of the Dirichlet-Neumann operator in 
this physical regime, we also derive a new generalization of the Matsuno
equations for $3D$ flows with uneven bottom.

\subsection{Asymptotic expansion of the Dirichlet-Neumann operator}

When $\eps$ and $\beta$ are small, it is possible to make
a Taylor expansion of 
$\cG_\mu[\eps\zeta,\beta b]\psi$ with respect to $\eps$ and $\beta$;
such an expansion has been derived for one dimensional surfaces in 
\cite{CGNS}; this method could be generalized to two dimensional surfaces,
but we chose here to use another technique based
on the following formulas
\begin{eqnarray*}
	\lim_{\eps\to 0}\frac{1}{\eps}\big(
	\cG_\mu[\eps\zeta,\beta b]\psi-
	\cG_\mu[0,\beta b]\psi\big)
	&=&-\cG_\mu[0,\beta b](\zeta \cG_\mu[0,\beta b]\psi)
	-\mu\nabla\cdot (\zeta(\nabla\psi))\\
	\label{formula}
	\lim_{\beta\to 0}\frac{1}{\beta}\big(
	\cG_\mu[0,\beta b]\psi-
	\cG_\mu[0,0]\psi\big)
	&=&{\mu}
	 \sech(\sqrt{\mu}\vert D\vert)\big[\nabla\cdot
	(b(\sech(\sqrt{\mu}\vert D\vert)\nabla\psi))\big],
\end{eqnarray*}
where we used the Fourier multiplier notation: given two functions
$f$ and $u$,  and denoting by $\widehat{\,}$  the Fourier transform,
$f(D)u$ is defined as:
$$
	\forall \xi\in \R^2, \qquad
	\widehat{f(D)u}(\xi):=f(\xi)\widehat{u}(\xi).
$$
We only prove the second of these formulas because the first one
can be established with the same techniques and can also be found in 
the literature 
(e.g. \cite{CShS}, and also Th. 3.20 of \cite{LannesJAMS} and 
Th. 3.1 of \cite{alvarezlannes} where a similar formula is established
for non necessarily flat surfaces).\\
First recall that by definition, $\cG_\mu[0,\beta b]\psi
=\dz\Phi[\beta b]_{\vert_{z=0}}$, where $\Phi[\beta b]$ solves
\begin{equation}\label{raplap}
	\left\lbrace
	\begin{array}{l}
	\mu\Delta\Phi[\beta b]+\dz^2\Phi[\beta b]=0, 
	\quad\mbox{ in }\quad -1+\beta b<z<0,\\
	\Phi[\beta b]_{\vert_{z=0}}=\psi,\qquad
	(\dz\Phi[\beta b]-\mu\beta\nabla\Phi[\beta b]\cdot\nabla b )_{\vert_{z=-1+\beta b}}=0
	\end{array}\right.
\end{equation}
(we denoted $\Phi[\beta b]$ instead of $\Phi$ to enhance the dependance
on the bottom topography). We thus have, for all smooth function
$\varphi$ compactly supported in $\overline{\Omega}$ (with $\overline{\Omega}=\{(X,z), -1+\beta b(X)\leq z\leq 0\}$), 
\begin{eqnarray*}
	\int_{\Omega} (\mu\Delta\Phi[\beta b]+\dz^2\Phi[\beta b])\varphi dXdz
	&+&\int_{z=0}(\Phi[\beta b]-\psi)\varphi dX\\
	&+&
	\int_{z=-1+\beta b}(\dz\Phi[\beta b]-
	\beta\mu\nabla b\cdot\Phi[\beta b])
	\varphi dX=0.
\end{eqnarray*}
Denoting $\chi=\lim_{\beta \to 0}\frac{1}{\beta}(\Phi[\beta b]-\Phi[0])$,
one can differentiate the above identity to obtain
\begin{eqnarray*}
	\int_{\Omega} (\mu\Delta\chi+\dz^2\chi)\varphi dXdz
	&+&\int_{z=0}\chi\varphi dX\\
	&+&
	\int_{z=-1}(\dz\chi-\mu\nabla b\cdot \nabla\Phi[0]+b\dz^2\Phi[0])
	\varphi dX=0.
\end{eqnarray*}
Since this identity holds for all test function $\varphi$, and since
$\dz^2\Phi[0]=-\mu\Delta\Phi[0]$, we deduce
that $\chi$ solves the boundary value problem
$$
	\left\lbrace
	\begin{array}{l}
	\mu\Delta\chi+\dz^2\chi=0,\quad\mbox{ in }\quad -1<z<0\\
	\chi_{\vert_{z=0}}=0,\qquad
	\dz \chi_{\vert_{z=-1}}=\mu\nabla\cdot( b\nabla\Phi[0]_{\vert_{z=-1}}).
	\end{array}\right.
$$
This problem can be solved explicitly:
$$
	\chi(\cdot,z)
	=\sqrt{\mu}
	\frac{\sinh(\sqrt{\mu }z\vert D\vert)}{\cosh(\sqrt{\mu}\vert D\vert)}
	\frac{\nabla}{\vert D\vert}\cdot (b\nabla\Phi[0]_{\vert_{z=-1}} );
$$
since moreover one has 
$\lim_{\beta\to 0}\frac{1}{\beta}(\cG_\mu[0,\beta b]\psi-\cG_\mu[0,0]\psi)
=\dz \chi_{\vert_{z=0}}$ and that $\Phi[0](\cdot,z)=\frac{\cosh(\sqrt{\mu}(z+1)\vert D\vert)}{\cosh(\sqrt{\mu}\vert D\vert)}\psi$, we get the formula.\\

A first order Taylor 
expansion of $\cG_\mu[\eps\zeta,\beta b]\psi$ with respect to 
$\eps$, together with the first formula, shows therefore that
\begin{equation}\label{exp1}
	\cG_\mu[\eps\zeta,\beta b]\psi=\cG_\mu[0,\beta b]\psi
	-\eps \cG_\mu[0,\beta b](\zeta \cG_\mu[0,\beta b]\psi)
	-\eps\mu\nabla\cdot(\zeta\nabla\psi)+O(\sqrt{\mu}(\eps\sqrt{\mu})^2);
\end{equation}
a first order Taylor expansion of $\cG_\mu[0,\beta b]\psi$ 
with respect to $b$, together with the second formula, gives also
\begin{equation}\label{exp2}
	\cG_\mu[0,\beta b]\psi=\cG_\mu[0,0]\psi+
	\mu\beta
	\sech(\sqrt{\mu}\vert D\vert)\big[
	\nabla\cdot (b(\sech(\sqrt{\mu}\vert D\vert)\nabla\psi))\big]+O(\sqrt{\mu}(\beta\sqrt{\mu})^2).
\end{equation}
Let us now define the operator ${\bf T}_\mu$ and $B_\mu$ as
\begin{equation}\label{defi}
	{\bf T}_\mu=-\frac{\tanh(\sqrt{\mu}\vert D\vert)}{\vert D\vert}\nabla
	\quad\mbox{ and }\quad
	{B}_\mu=\sech(\sqrt{\mu}\vert D\vert)\big[
	b(\sech(\sqrt{\mu}\vert D\vert)\cdot)\big]
\end{equation}
We thus have $\cG_\mu[0,0]\psi=\sqrt{\mu}{\bf T}_\mu\cdot\nabla\psi$, and
(\ref{exp1}) and (\ref{exp2}) show that
\begin{eqnarray}
	\nonumber
	\cG_\mu[\eps\zeta,\beta b]\psi&=&
	\sqrt{\mu}{\bf T}_\mu\cdot\nabla\psi+
		\mu\beta\nabla\cdot (B_\mu\nabla\psi)
	-\eps\mu {\bf T}_\mu\cdot\nabla(\zeta  
	{\bf T}_\mu\cdot\nabla\psi)\\
		\label{formulefinale}
	& &-\eps\mu\nabla\cdot(\zeta\nabla\psi)
	+ O(\sqrt{\mu}(\eps\sqrt{\mu})^2,
	\sqrt{\mu}(\beta\sqrt{\mu})^2).
\end{eqnarray}
\begin{rema}
	In the shallow water, weakly nonlinear regime, and with a weakly
	nonlinear topography (that is, $\mu\ll1$, $\eps=O(\mu)$, 
	$\beta=O(\mu)$), one can deduce from (\ref{formulefinale}) that
	$$
		\cG_\mu[\eps\zeta,\beta b]\psi=
		-\mu \nabla\cdot (h\nabla \psi)
		-\frac{\mu^2}{3}\nabla\cdot \Delta\nabla\psi+O(\mu^3),
	$$
	and we thus recover the approximation derived with another
	technique in \S \ref{sectweak} (see Remark \ref{remrec}).
\end{rema}

\subsection{Derivation of a fully dispersive model for 
$3D$ flows over uneven bottoms}

We derive here a new system of fully dispersive equations which generalizes
to the case of $2D$ surfaces and nonflat bottoms the systems
derived by Matsuno ($1D$ surfaces, flat \cite{matsuno1} and uneven \cite{matsuno2} bottoms), Choi \cite{choi} and Smith \cite{smith} 
($2D$ surfaces, flat bottoms).\\
We first define the horizontal velocity at the surface as 
${\bf v}_S=(\nabla\Phi)_{\vert_{z=\eps \zeta}}$, where $\Phi$
is the velocity potential given by (\ref{raplap}). By definition of 
$\psi$ and $\cG_\mu[\eps\zeta,\beta b]\psi$, we get
\begin{eqnarray*}
	\nabla\psi&=&{\bf v}_S+\eps\nabla\zeta(\dz\Phi)_{\vert_{z=\eps\zeta}}\\
	&=&{\bf v}_S+\eps\frac{\cG_\mu[\eps\zeta,\beta b]\psi+\eps\mu\nabla\zeta\cdot\nabla\psi}{1+\eps\mu\vert\nabla\zeta\vert^2}\nabla\zeta\\
	&=&{\bf v}_S+\eps\sqrt{\mu}({\bf T}_\mu\cdot{\bf v}_S)\nabla\zeta+O((\eps\sqrt{\mu})^2),
\end{eqnarray*}
where we used (\ref{formulefinale}) and $\nabla\psi={\bf v_S}+O(\eps\sqrt{\mu})$ for the last relation. Plugging this relation into (\ref{formulefinale}),
one gets similarly
\begin{eqnarray}
	\nonumber
	\frac{1}{\sqrt{\mu}}\cG_\mu[\eps\zeta,\beta b]\psi&=&
	{\bf T}_\mu\cdot{\bf v}_S+
	\sqrt{\mu}\beta\nabla\cdot (B_\mu{\bf v}_S)
	-\eps\sqrt{\mu} {\bf T}_\mu\cdot(\zeta  
	\nabla {\bf T}_\mu\cdot{\bf v}_S)\\
	\label{expDNdeep}
	& &-\eps\sqrt{\mu}\nabla\cdot(\zeta{\bf v}_S)
	+ O(\sqrt{\mu}(\eps\sqrt{\mu})^2,
	\sqrt{\mu}(\beta\sqrt{\mu})^2).
\end{eqnarray}
Taking the gradient of the second equation 
of (\ref{nondimww}) and 
using the above two identities gives therefore the following
set of deep-water equations:
\begin{equation}\label{deep}
	\left\lbrace
	\begin{array}{l}
	\displaystyle \dt \zeta-\frac{1}{\sqrt{\mu}\nu}
	{\bf T_\mu}\cdot {\bf v}_S
	+\frac{\eps\sqrt{\mu}}{\sqrt{\mu}\nu}
	\big({\bf T}_\mu\cdot(\zeta\nabla{\bf T}_\mu\cdot{\bf v}_s)
	+\nabla\cdot(\zeta{\bf v}_S)\big)\\
	\displaystyle
	\indent\indent\indent\indent\indent\indent\indent\indent	
	=\frac{\beta\sqrt{\mu}}{\sqrt{\mu}\nu}\nabla\cdot(B_\mu{\bf v}_S)
	\vspace{0.5mm}\\
	\displaystyle \dt {\bf v}_S+\nabla\zeta+\eps\sqrt{\mu}
	\big(\frac{1}{2\sqrt{\mu}\nu}\nabla\vert{\bf v}_S\vert^2-\nabla\zeta{\bf T}_\mu\cdot\nabla\zeta\big)=0,
	\end{array}\right.
\end{equation}
where we recall that ${\bf T}_\mu$ and $B_\mu$ are defined in (\ref{defi}),
and that $\nu=\tanh(\sqrt{\mu})/\sqrt{\mu}$ (so that one can replace 
$\sqrt{\mu}\nu$ by $1$ in (\ref{deep}) in deep water).
\begin{rema}
	If we remove the topography term $B_\mu$ from these 
	equations, we recover the two-dimensional equations
	(3.25)-(3.26) derived by Choi \cite{choi}. 
	If we take the one dimensional
	version of (\ref{deep}) we recover the equations derived by Matsuno
	(Eqs (19)-(20) of \cite{matsuno1} for flat bottoms
	and (4.28)-(4.29) of \cite{matsuno3} for nonflat bottoms).
\end{rema}
\begin{rema}
	The equations (\ref{deep}) are precise up to order 
	$O(\eps\sqrt{\mu},\eps\sqrt{\beta})$
	in deep water (since $\nu\sim \mu^{-1/2}$); one could
	also use them in shallow water, but they are then precise
	up to order $O(\eps,\beta)$ only (since $\nu\sim 1$ in shallow water).
	This is the same accuracy as the one provided by
	the weakly nonlinear shallow water models
	(with a weakly nonlinear topography) of Section \ref{sectweak};
	it is therefore not surprising to check that
	one recovers the Boussinesq system (\ref{weakBouss})
	from (\ref{deep}) by a simple Taylor expansion of the
	operators ${\bf T}_\mu$ and $B_\mu$ and by observing
	that ${\bf v}=(1+\mu\frac{1}{3}\Delta){\bf v_s}+O(\mu^2)$.\\
	In the fully nonlinear regime (with fully nonlinear topography)
	studied in Section \ref{sectSerre}, the equations (\ref{deep})
	are precise up to order $O(\eps,\beta)=O(1)$ while we
	saw that the Green-Naghdi equations (\ref{eqGN}) are precise
	up to order $O(\mu)$. It is 
	therefore not surprising to check that the shallow water limit
	of (\ref{deep}) does NOT give the Green-Naghdi equations
	(\ref{eqGN}).
\end{rema}
\begin{rema}
The system (\ref{deep}) is ``fully dispersive'' in the sense that
its dispersion relation is the same as for the full water-wave 
equations. This is the case because the expansion (\ref{expDNdeep})
keeps the nonlocal effects of the Dirichlet-Neumann operator
$G_\mu[\eps\zeta,\beta b]\psi$. In \cite{AMS,MBS,MBL}, the authors make
a differential approximation of shallow water type based on Pad\'e
approximants; they show that the dispersive properties remain good
far beyond the shallow water regime when bathymetric changes are not
too strong (in this latter case, the model is more complex and
its range of validity much narrower \cite{MFW}).

\end{rema}

%
%
\newpage
\section{Conclusion}

In this paper we have presented a systematic derivation of the main 2D asymptotic models for shallow and deep water waves, which allows to clarify their validity domain. We have also derived a new 2D fully dispersive model, system (\ref{deep}), for small wave steepness which extends to uneven bottom the approach developed by Matsuno \cite{matsuno3} and Choi \cite{choi}. 
We have shown that even though these models remain valid in shallow water, 
their precision is then far below what can be achieved with
the Green-Naghdi or Serre model when fully nonlinear 
waves are considered ($\eps\sim 1$).
Hence, contrary to what it is generally thought \cite{matsuno1,matsuno2,matsuno3,choi} this approach cannot in practical be considered as
a unified theory of nonlinear waves because it is not accurate enough for 
nonlinear shallow water waves ($\epsilon \sim 1$, $\mu \ll 1$). 
In particular, we have shown that system (\ref{deep}) in the shallow water 
limit ($\mu \ll 1$) does not correspond to the correct fully nonlinear 
equations, namely the Green-Naghdi or Serre equations. 
Another reason why these fully dispersive water models are not
likely to furnish interesting models in shallow water is that 
there is no obvious shoreline boundary condition for them. 

The Green-Naghdi equations represent the appropriate model to describe nonlinear shallow water wave propagation and wave oscillations at the shoreline. For coastal applications, the Green-Naghdi equations can be easily extended to include accurate linear dispersive effects, which allow to describe shoaling processes in intermediate water depth (see Wei et al. \cite{{wei1995}} and Cienfuegos et al. \cite{cie2006,cie2007}). Another interest of these equations is that
there is a natural shoreline boundary condition given by the flux conservation
equation (the flux $h{\bf v}$ vanishes at the
shoreline in the first equation of \ref{eqGN}).
However, for deep water wave propagation the Green-Naghdi model is no more 
valid and the fully dispersive model (\ref{deep}) must therefore 
be used. 

It follows from these considerations that the asymptotic description
of coastal flows requires at least the use of 2 different models: one
for shallow-water (e.g. the Green-Naghdi equations (\ref{eqGN})) and
another one for deep water (e.g. (\ref{deep})). The numerical
coupling of these two models is therefore a natural perspective for
further works.

\vskip1cm
\noindent
{\bf ACKNOWLEDGMENT}

This work has been supported by ACI Jeunes Chercheurs "Dispersion et nonlinéarités" and was also performed within the framework of the LEFE-IDAO program (Interactions et Dynamique de l'Atmosphère et de l'Océan) sponsored by the CNRS/INSU.

\end{document}